\documentstyle[epsf,epsfig,here,12pt]{article}   
\begin{document} 
\newcommand{\BK}{B_K} 
\newcommand{\Vcb}{\left | {\rm V}_{cb} \right |} 
\newcommand{\Vub}{\left | {\rm V}_{ub} \right |} 
\newcommand{\VubVcb}{\left | {\rm V}_{ub} / {\rm V}_{cb} \right |} 
\newcommand{\rhobar}{\overline{\rho}} 
\newcommand{\etabar}{\overline{\eta}} 
\newcommand{\epsilonk}{\left|\varepsilon_K \right|} 
\newcommand{\vubovcb}{\left | \frac{V_{ub}}{V_{cb}} \right |} 
\newcommand{\vubsvcb}{\left | V_{ub}/V_{cb}  \right |} 
\newcommand{\vtdovts}{\left | \frac{V_{td}}{V_{ts}} \right |} 
\newcommand{\epsp}{\frac{\varepsilon^{'}}{\varepsilon}} 
\newcommand{\dmd}{\Delta m_d} 
\newcommand{\dms}{\Delta m_s} 
\newcommand{\Bs}{\rm{B^0_s}} 
\newcommand{\Bsb}{\overline{\rm{B}^0_s}} 
\newcommand{\Bp}{\rm{B^{+}}} 
\newcommand{\Bm}{\rm{B^{-}}} 
\newcommand{\Bo}{\rm{B^{0}}} 
\newcommand{\Bd}{\rm{B^{0}_{d}}} 
\newcommand{\Bdb}{\overline{\rm{B}^{0}_{d}}} 
\newcommand{\Dstarp}{\mbox{D}^{\ast +}} 
\newcommand{\Dstarm}{\mbox{D}^{\ast -}} 
\newcommand{\Dsstarp}{\mbox{D}_s^{\ast +}} 
\newcommand{\GeV}{\rm{GeV}} 
\newcommand{\MeV}{\rm{MeV}} 
\newcommand{\Dstar}{{\rm D}^{\ast}} 
\newcommand{\fbdsqbd}{f_{B_d} \sqrt{\hat B_{B_d}}} 
\newcommand{\fbssqbs}{f_{B_s} \sqrt{\hat B_{B_s}}} 
\begin{titlepage} 
 
\Large 
\centerline {\bf Determination of Unitarity Triangle parameters:} 
\centerline {\bf    where do we stand ? } 
\normalsize 
  
\vskip 2.0cm 
\centerline {Achille Stocchi~\footnote{email: Stocchi@lal.in2p3.fr} 
\footnote{work done in collaboration with M.~Ciuchini, G.~D'Agostini,  
E.~Franco, V.~Lubicz, G. Martinelli, F. Parodi and P. Roudeau}} 
\centerline {LAL et Universit\'e de Paris-Sud, BP 34, F91898 - Orsay, France} 
\vskip 1.0cm 
  
\begin{abstract} 
In this note I review the current determination of the unitarity  
triangle parameters by using the theoretical and  
experimental information available in summer 2000. 
\end{abstract} 
\end{titlepage} 
\newpage 
 
\section{Prologo} 
\begin{figure} 
{\epsfig{figure=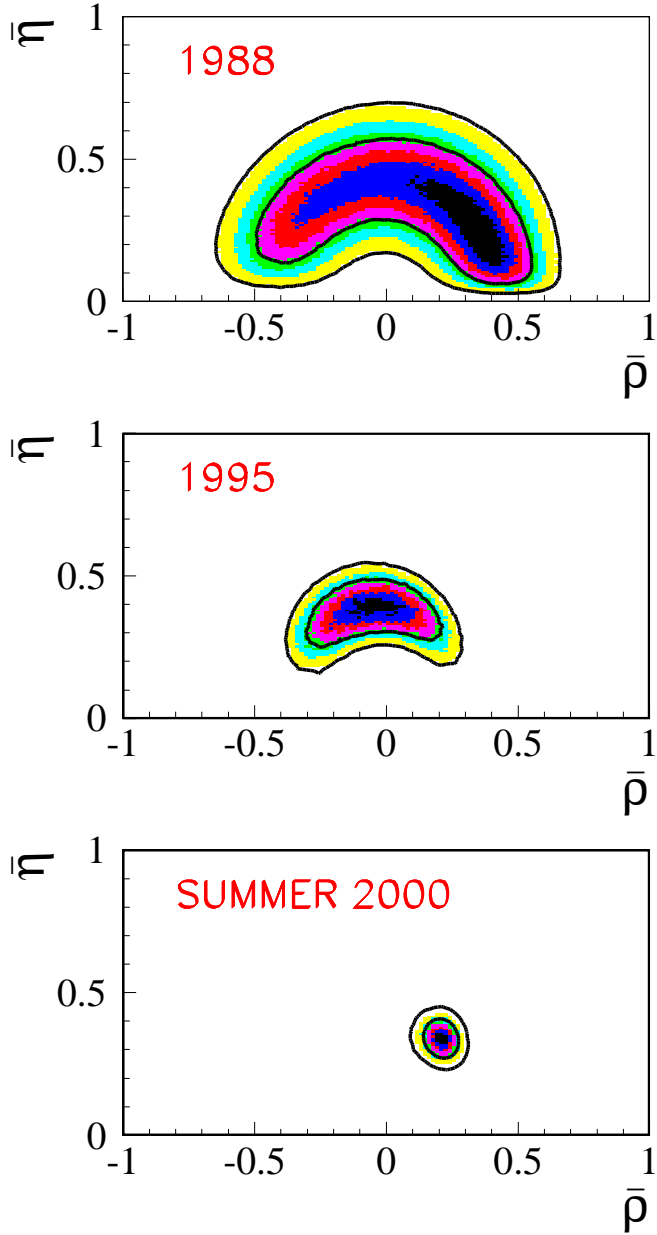,bbllx=10pt, 
bburx=330pt,bblly=10pt,bbury=530pt,height=16cm}} 
\caption[]{\it {The allowed region for $\rhobar$ and  
$\etabar$ (in 1988, 1995 and summer 2000) 
using the constraints given by the measurements of  
$\left | V_{ub} \right |/\left | V_{cb} \right |$,  
$\epsilonk$, $\Delta m_d$ and $\Delta m_s$.  
The contours at 68 $\%$ and 95 $\%$ probability are shown.}} 
\label{fig:fig1} 
\end{figure}

Tremendous improvements in the determination of the unitarity triangle  
parameters have been achieved during the last 10 years as illustrated by the  
reduction of the selected region in the ($\rhobar-\etabar$) plane shown in  
Figure \ref{fig:fig1}.  
What are the major developments responsible for this success ? 
 
{\bf * (a)} The continuous and precious work done by the CLEO Collaboration; 
 
{\bf * (b)} The precise and somehow ``unexpected'' results  
obtained by the SLD and LEP Collaborations,  
which provided the main contribution in the development from the 1995 to  
2000 configurations shown in Fig.~\ref{fig:fig1}; 
 
{\bf * (c)} The top quark discovery and the accurate measurement of  
its mass at the TeVatron; 
 
{\bf * (d)} The improvements in Lattice QCD calculations; 
 
{\bf * (e)} The improvements in the theoretical calculations used in extracting  
$\Vcb$ and $\Vub$. 
 
In this short note, I summarize the results of Ref.~\cite{ref:osaka2000}, where 
we determine the parameters of the unitarity triangle using all available  
recent measurements and theoretical calculations (mainly lattice QCD). 
Further details can be found in Ref.~\cite{ref:previous1}. 
I also attempt to demonstrate the robustness of our results. 
 
\section{The main actors (Allegro con brio, crescendo continuo)} 
\label{sec:sec2} 
 
The central values and the uncertainties for the relevant input  
parameters used in this analysis are given in Table~\ref{tab:1}. 
In the following I give short comments on the determination of the different  
parameters: 
 
{\boldmath $\Vcb$} -- Two methods are used to extract $\Vcb$.  
The first one makes use of the inclusive 
semileptonic decays of B-hadrons and the theoretical calculations to extract  
$\Vcb$ are done in the framework of the Operator-Product-Expansion (OPE). 
A second method uses the exclusive  
decays, $\Bdb \rightarrow \Dstarp \ell^- \overline{\nu_{\ell}}$.  
In this case, the value of $\Vcb$ is obtained by measuring the differential  
decay rate at maximum mass of the  
charged lepton-neutrino system, $q^2$, in the framework of HQET. 
 
The present exclusive measurements are marginally compatible  
(the fit probability is 6$\%$). 
A procedure \cite{ref:dago1}, developed to best combine results which  
may be in mutual disagreement, has been used to determine 
the quoted central value for $\Vcb$ shown in Table~\ref{tab:1} 
using inclusive (LEP) and exclusive (LEP and CLEO) measurements. 
We regard this value as the present world average. 
 
{\boldmath $\VubVcb$} -- The CLEO collaboration \cite{ref:vubcleo} has measured  
the branching fraction for the decay, 
$\overline{{\rm B_d^0}} \rightarrow \rho^+ \ell^- \overline{\nu_{\ell}}$. 
The value of $\Vub$ is then deduced using several models. 
The LEP experiments \cite{ref:vublep} have measured  
$Br(b \rightarrow u \ell \bar{\nu}_l X)$ with less statistical precision  
than does CLEO, but with reduced systematic uncertainties. 
In events with an identified high transverse momentum lepton,  
they use several kinematical variables, which allow discrimination  
between b $\rightarrow$ c and b $\rightarrow$ u transitions.  
Using models based on the OPE, a value for $\Vub$ is then obtained.  
These two measurements are shown in Table~\ref{tab:1}.  
The uncertainties are uncorrelated between CLEO and LEP results. 
 
{\boldmath $\dms$} -- After the addition of recent measurements from the 
SLD/LEP collaborations, the limit on $\Delta \rm{m}_s$, at 95$\%$ C.L.  
has not increased much compared to last year's result. 
But, the  
sensitivity has improved a lot reaching $\Delta m_s = 18.0 \, ps^{-1}$  
\cite{ref:stocchi} 
(the sensitivity corresponds to the value of $\dms$ ($\Delta m_s^{sens}$) at which  
it is expected that the 95$\%$ limit will be set in 50\% of the ideal experiments  
having the same caracteristics as the real data, if the true value of  $\dms$ is much  
larger than $\Delta m_s^{sens}$ ) 
 
{\bf Non-perturbative QCD parameters} -- In the framework of lattice QCD, 
important improvements have been recently achieved in the evaluation of  
non-perturbative QCD parameters for B hadrons.  
As a consequence, we use only the most recent values. 
The central values and the uncertainties given in Table~\ref{tab:1}  
have been evaluated in Ref.~\cite{ref:osaka2000}  
and are in good agreement with those given in the three 
most recent reviews in Ref.~\cite{ref:lattice1}. 
 
\section{The results (Andante allegro)} 
The region in the ($\rhobar,~\etabar)$ plane selected by the measurements of  
$\epsilonk$, $\VubVcb$, $\dmd$ and from the limit on $\dms$, is shown at  
the bottom of Fig.~\ref{fig:fig1}. Our fit values for $\rhobar$, $\etabar$, 
and the angles are given in Table~\ref{tab:stab}. 
  
Our value, $\sin(2\beta) = 0.698 \pm 0.066$, is rather precisely determined, 
compared with the world average, $\sin(2\beta) = 0.52 \pm 0.22$, measured  
using $B \rightarrow J/\psi K_S$ events. 
The angle $\gamma$ is known within an accuracy of about 10$\%$.  
The probability that $\gamma$ is greater  
than 90$^{\circ}$ is only 0.03$\%$. This result is mainly due to  
the improved sensitivity on $\dms$  
and is very slightly dependent on the value and on the error assigned  
to the $\xi$ parameter. ( $\xi={f_{B_s}\sqrt{\hat {B_{B_s}}}}/{ f_{B_d}\sqrt{\hat {B_{B_d}}}}$ 
as indicated in Table~\ref{tab:1}). 
The central value for the angle $\gamma$ is more than 2$\sigma$ smaller 
than that obtained in recent fits of rare $B$-meson  
two-body decays~\cite{ref:cleorare} (see for istance \cite{ref:cleorare}-d,  
where $\gamma = (114 ^{+24}_{-23})^{\circ}$)  
 
\begin{figure} 
\begin{center} 
{\epsfig{figure=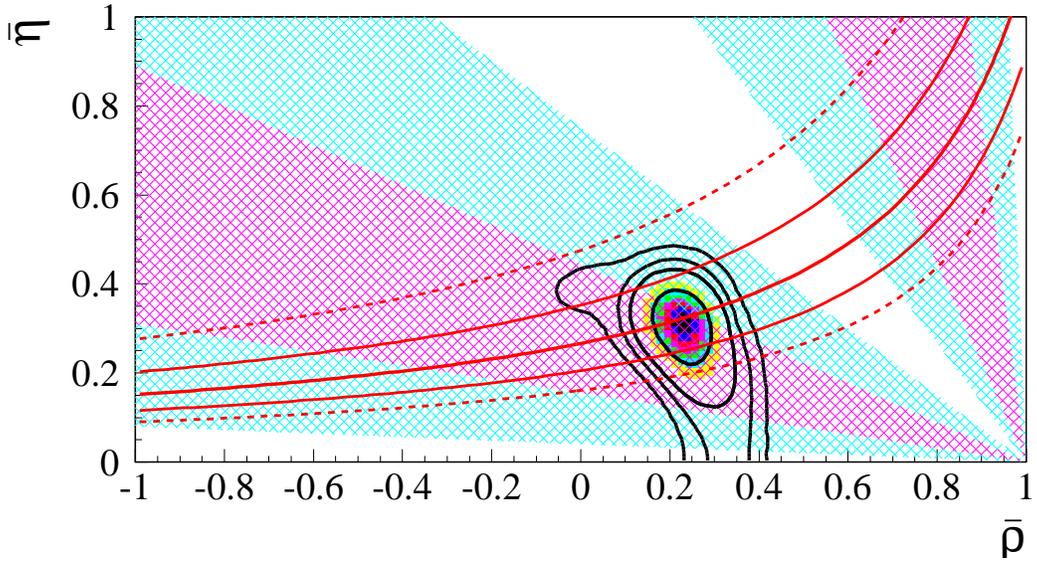,bbllx=30pt,bburx=503pt,bblly=1pt,bbury=270pt,height=8cm}} 
\caption{ \it{The allowed regions (at 68\%, 95\%, 99\% and 99.9\% probability) for $\rhobar$ and 
$\etabar$ using  the constraints given by the measurements of  $\left | V_{ub} \right |/\left | V_{cb} 
\right |$,  $\Delta m_d$ and  $\Delta m_s$. The constraint due to $\epsilonk$ is not included. The regions 
(at 68\% and 95\% probability) selected by the measurements of $\epsilonk$ (continuous (1$\sigma$) and dotted (2$\sigma$) curves) 
and $\sin(2\,\beta)$ (darker (1$\sigma$) and clearer (2$\sigma$) zones) are shown.  
For $\sin(2\,\beta)$ the two solutions are displayed.}} 
\label{fig:noepk} 
\end{center} 
\end{figure}

An interesting study consists of removing the theoretical constraint for  
$\hat \BK$ in the measurement of $\epsilonk$.  
The corresponding selected region in the ($\rhobar,~\etabar)$  
plane is shown in  Figure~\ref{fig:noepk}. 
In Figure~\ref{fig:noepk} the regions (at 68\% and 95\% probability) selected by the  
measurement of $\epsilonk$ alone are also drawn.  
This comparison shows that the Standard Model picture of CP violation in the $K$ system and 
of $B$ decays and oscillations is consistent.  
This constitutes already a test of compatibility  
between the measurements of the sides and of the angles of the CKM triangle. 
This can be quantified by comparing the value, $\hat \BK = 0.87 \pm 0.06 \pm 0.13$,  
obtained from  
lattice-QCD calculations,  with the one extracted by using constraints from  
$b$-physics alone, 
$\hat \BK = 0.90^{+0.30}_{-0.14}$. In the same figure, we also compare  
the allowed regions  
in the ($\rhobar,~\etabar)$ plane with those selected by the  
measurement of $\sin(2\,\beta)$ using  
$J/\psi K_S$ events. \\ 
It is informative to remove other theoretical or experimental constraints from a  
fit (as $\hat \BK$ in the 
previous example). This illustrates the effectiveness of a  
constraint and gives a most like value of a parameter  
within the Standard Model. The most significant results we find are : 
\begin{equation} 
\fbdsqbd      =  (230 \pm 12)~\MeV\, ,\quad \dms           
=  (16.3 \pm 3.4) ps^{-1} 
\label{eq:dms} 
\end{equation} 
 
\section{Stability Tests (Adagio..... con calma)} 
 
To determine the robustness of our results, we investigate how the quoted  
accuracies on the unitarity triangle parameters change if: 
(a) the errors on the input parameters are changed, or  
(b) a different statistical method is used to obtain the results. 
 
{\bf (a)} -- It is a  
basic exercise to verify how the accuracy quoted on final results 
depends on the assumed errors for the different input parameters.  
A similar analysis has been already presented in \cite{ref:previous1}. 
As shown in Table~\ref{tab:stab}, all the flat theoretical errors, 
as well as the error on $\Vcb$, are multiplied by a factor of 2. 
Thus, the assumed error on $\Vcb$ is ($\pm 3.8\times 10^{-3}$);  
on  $\hat \BK$, it is ($\pm$ 0.06(Gaussian) $\pm$ 0.26(flat)). 
The main conclusion of this study is that, even in this extreme 
case, the unitarity triangle parameters are determined 
with uncertainties which increase by about a factor of 1.6. 
 
\begin{figure}[htb!] 
\begin{center} 
\epsfig{file=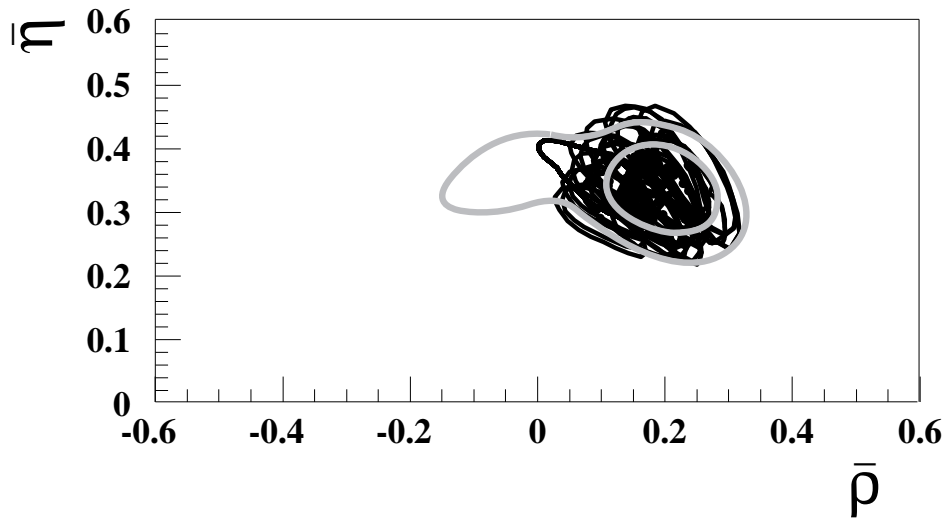,width=13cm}  
\caption[]{\it {The contour corresponding to 95$\%$ probability,  
using our approach (continous contour), is compared with the envelope of  
95$\%$ C.L. ellipses, obtained using 
the 95$\%$ C.L. ``scanning method''.  
In the former case, the 68$\%$ contour is also shown.  
The parameters used in this study are those given in Ref.~\cite{ref:plasz}.  
The allowed region in the ($\rhobar$-$\etabar$) plane is larger than the one shown  
at the bottom of Fig.~\ref{fig:fig1}  
because input parameters correspond,respectively, to  
the current knowledge at the beginning of 1998 and summer 2000}} 
\label{fig:scanning standard} 
\end{center} 
\end{figure}

2) A comparison between our results (using a Bayesian approach, see Ref.  
\cite{ref:osaka2000} for more details) and those obtained using the ``scanning method''   
(adopted by the Babar Collaboration), has been done. 
 To do this comparison, the same central  
values and errors for the parameters have been used in the two cases. 
 When a parameter is scanned, in the ``scanning approach'', a flat distribution  
corresponding to the scanning range 
 is used in our approach. 
These parameters are those given in ~\cite{ref:plasz}.  
The ``95$\%$ C.L.'' contours obtained with the two methods  
are compared in Fig.~\ref{fig:scanning standard}. 
The main conclusion is that, when the same input parameters  
are used, very similar results are obtained using the two methods.  
We therefore do not believe that our method yields ``optimistic'' results. 
 
\section{Conclusions (Finale con brio)} 
The determination of the unitarity triangle parameters has  
already entered in a mature age, the age of precision tests.  
I have illustrated the impressive improvements on the  
determination of the two sides of the  
unitarity triangle using only B decays and oscillations. 
Our results are shown to be robust and stable against changes in the 
uncertainties of the input parameters and against the statistical method 
used to obtain them. 
 
The selected region in the ($\rhobar$-$\etabar$) plane is  
compatible with the measurement  
of CP violation in the Kaon system. Similar tests are 
expected soon from the direct measurement of 
sin(2$\beta$) at B-Factories and future hadron machines. 
 
\section*{Acknowledgements} 
Thanks to the organizers of BEAUTY 2000 for the impeccable 
organization. A special thank to Yoram Rozen who  
really made the week spent in Israel unforgettable and set  
a new standard for organizing conferences. 
Thanks to Patrick Roudeau and Peter Schlein for the careful  
reading of this document.

\pagebreak 
 
\begin{table}[htb!] 
\begin{center} 
\frame{ 
\footnotesize{ 
\begin{tabular}{c|c|c|c|c} 
\hline 
        Parameter                  &           Value           
& Gaussian ($\sigma$)  &  Uniform (half-width)  & Ref. \\  
\hline  
$\left | V_{cb} \right |$          &    $41.0 \times 10^{-3}$  
& $1.6\times 10^{-3}$  & --      & \cite{ref:osaka2000} \\ 
$\left | V_{ub} \right |$ (CLEO)   & $3.25\times 10^{-3}$      
& $0.29\times 10^{-3}$ & $0.55\times 10^{-3}$ &\cite{ref:vubcleo}  \\ 
$\left | V_{ub} \right |$ (LEP)       & $4.04\times 10^{-3}$  
& $0.63\times 10^{-3}$ & $0.31\times 10^{-3}$ & \cite{ref:vublep}   \\ 
     $\Delta m_d$  & $0.487~\mbox{ps}^{-1}$ & $0.014~\mbox{ps}^{-1}$  
& --        &~\cite{ref:stocchi}  \\ 
$\Delta m_s$  & $>$ 15.0 ps$^{-1}$   & \multicolumn{2}{|c|}{see text}  
&~\cite{ref:stocchi}\\ 
$m_t$ & $167~\GeV$ & $ 5~\GeV$ & --   &~\cite{ref:top} \\ 
$\hat \BK$    & $0.87$ & $0.06$ &  $0.13$       &  \cite{ref:osaka2000}  \\ 
$f_{B_d} \sqrt{\hat {B_{B_d}}}$ & $230~\MeV$  & $25~\MeV$       
&  $20~\MeV$  & \cite{ref:osaka2000} \\ 
$\xi=\frac{ f_{B_s}\sqrt{\hat {B_{B_s}}}}{ f_{B_d}\sqrt{\hat {B_{B_d}}}}$       
& $1.14$ & 0.04 & $ 0.05 $& 
 \cite{ref:osaka2000} \\ 
\hline  
\end{tabular} }}  
\caption[]{ \it {Values of the quantities entering into the  
expressions of $\epsilonk$, $\VubVcb$, $\dmd$ and $\dms$. 
The Gaussian and the flat part of the error are given explicitly. }} 
\label{tab:1}  
\end{center} 
\end{table} 
 
\begin{table}[htb!] 
\begin{center} 
\begin{tabular}{|c|c|c|c|} 
\hline 
    Parameters  &      Std. Result              
&     Theo. x 2~,~ $\Vcb$ x 2              
&   Maximal Increase     \\ 
\hline 
    $\rhobar$    &     0.224 $\pm$ 0.038        
&   $\pm$ 0.064             &   $\sim$1.7     \\ 
    $\etabar$    &     0.317 $\pm$ 0.040        
&   $\pm$ 0.065             &   $\sim$1.6     \\ 
    sin2$\beta$  &    0.698 $\pm$ 0.066       
&   $^{+0.086}_{-0.101}$    &   $\sim$1.4     \\ 
    sin2$\alpha$ &    -0.42 $\pm$ 0.23       
&   $\pm$ 0.37             &   $\sim$1.6     \\ 
    $\gamma$     &     (54.8 $\pm$ 6.2)$^o$     
&   $(\pm$ 10.0)$^o$    &   $\sim$1.6     \\ 
\hline 
\end{tabular} 
\caption[]{ \small {\it Stability tests. Variation of the error on some unitarity triangle parameters obtained 
by multiplying the flat part of the theoretical error by a factor 2 as well as the error on $\Vcb$ by the same  
factor.}} 
\label{tab:stab}  
\end{center} 
\end{table}  
 
\pagebreak 
 
\end{document}